\documentclass{kapproc} 

\usepackage{procps}

\usepackage[dvips]{graphicx}

\upperandlowercase

\kluwerbib 

\begin{document}

\articletitle{Star-Forming Complexes in Galaxies }


\author{Bruce G. Elmegreen}
\affil{IBM Research Division, T.J. Watson Research Center\\
1101 Kitchawan Road, Yorktown Hts., NY,
USA}\email{bge@watson.ibm.com}

\begin{abstract}

Star complexes are the largest globular regions of star formation
in galaxies. If there is a spiral density wave, nuclear ring,
tidal arm, or other well-defined stellar structure, then
gravitational instabilities in the gaseous component produce giant
cloud complexes with a spacing of about three times the width.
These gas complexes form star complexes, giving the familiar beads
on a string of star formation along spiral arms, or nuclear
hotspots in the case of a ring. Turbulence compression,
supernovae, and self-gravitational contraction inside the giant
clouds produce a nearly scale-free structure, including giant
molecular clouds that form OB associations and molecular cloud
cores that form clusters. Without stellar density waves or similar
structures, random gravitational instabilities form flocculent
spirals and these fragment into star complexes, OB associations
and star clusters in the same way. In this case, all of the
structure originates with gravitational instabilities and
turbulence compression, but the usual concept of a star complex
applies only to the largest globular object in the hierarchy,
which has a size defined by the flocculent arm width or galaxy
thickness. The largest coherent star-forming regions are the
flocculent arms themselves. At the core of the hierarchy are the
very dense clumps in which individual and binary stars form. The
overall star formation rate in a galaxy appears to be regulated by
gravitational collapse on large scales, giving the
Kennicutt/Schmidt law scaling with density, but the efficiency
factor in front of this scaling law depends on the fraction of the
gas that is in a dense form. Turbulence compression probably
contributes to this fraction, producing a universal efficiency on
galactic scales and the observed star formation rate in disk
systems. The CO version of the Schmidt law, recently derived by
Heyer et al. (2004), follows from the turbulent hierarchy as well,
as do the local efficiencies of star formation in OB associations
and clusters. The efficiency of star formation increases with
cloud density, and this is why most stars form in clusters that
are initially self-bound.
\end{abstract}


\noindent To be published in ``The many scales in the Universe -
JENAM 2004 Astrophysics Reviews,'' from the Joint European and
National Astronomical Meeting in Granada, Spain, September 13-17,
2004. Kluwer Academic Publishers, edited by Jose Carlos del Toro
Iniesta, et al.

\section*{Introduction to Star Complexes}
Star complexes are the largest coherent groupings of young stars
in galaxies. Shapley (1931) first noted that they appear like
``small irregular star clouds'' up to 400 pc in size in the Large
Magellanic Clouds. McKibben Nail \& Shapley (1953) later defined
12 ``constellations'' in the LMC. Baade (1963) considered the most
active examples of star complexes and called them
``superassociations,'' such as 30 Dor in the LMC and NGC 206 in
M31.

The definition of star complexes was broadened by Efremov (1979).
He considered complexes to be collections of Cepheid variables
with similar periods and velocities, that is, within a narrow
range of ages (e.g., $\sim50$ My) and moving as a group.  This
makes them distinct from ``OB associations,'' which are
collections of OB stars within a narrower range of ages ($\sim10$
My). Efremov noted that the OB associations in M31 defined by van
den Bergh (1981) have an average size of 480 pc, which is much
larger than local OB associations (80 pc).  He concluded that the
M31 associations are complexes, and that star complexes generally
contain OB associations as sub-parts.  Battinelli et al. (1996)
quantitatively found stellar groupings in M31 and demonstrated
this 2-component, or hierarchical, nature of associations and
complexes.

Hierarchical structure in the LMC stars was first quantified by
Feitzinger \& Galinski (1987).  Most recently, Maragoudaki et al.
(1998) measured stellar groupings in the LMC using discrete
magnitude limits in the U-band. They noted that the smaller
groupings in the hierarchy are rounder.  Gouliermis et al. (2000)
did the same for B stars in LMC fields. Hierarchical structure as
a general property of interstellar gas was discussed much earlier
than this (see review in Scalo 1985). The observation that stellar
groupings generally have the same type of hierarchical structure
as the gas is not surprising since the stars form from the gas
(e.g., see Bonnell, Bate \& Vine 2003).

The definition of star complexes was broadened further by
Elmegreen \& Efremov (1996) to mean the largest ``globular'' scale
in a hierarchy of star formation ranging from multiple stars to
flocculent spiral arms.  They showed that bigger scales evolve
slower, with $t\propto L^{0.5}$, that all scales evolve on about a
dynamical crossing time, and that the largest globular scale for
star formation should be about the disk thickness (times
$\sim\pi$), at which point the shear time becomes comparable to
the crossing time. Bigger regions form the same way and are part
of the hierarchy, but they look like flocculent spiral arms
instead of roundish star complexes. Efremov \& Elmegreen (1998)
also found that the size-duration correlation for star formation
is about the same as the size-crossing time correlation for
molecular clouds, which suggests that turbulence regulates star
formation on scales comparable to or smaller than the ambient ISM
Jeans length.

The observation that star formation is somewhat scale-free, even
up to $\sim0.1$ times galactic scales, received considerable
support after the gas was found to be scale-free over similar
lengths. Fractal structure in the gas on very small scales had
been observed for a long time (e.g., Falgarone, Phillips \& Walker
1991 and references therein), but the first observations of
fractal structure in whole galaxies was by Westpfahl et al. (1999)
and Stanimirovic et al. (1999).  Westpfahl et al. found the
fractal dimension for gas in M81 group galaxies using
area-perimeter relations and box counting techniques on HI maps,
while Stanimirovic et al. found that the power spectrum of HI
emission from the entire Small Magellanic Clouds is a scale-free
power law. A similar power-law was later found for the LMC
(Elmegreen, Kim \& Staveley-Smith 2001). In the LMC, much of the
gas also resembles shells rather than blobs, and these shells are
hierarchical too ({\it ibid.}). The exact relation between this
shell structure, turbulence, and star formation is not clear yet
(Wada, Spaans, \& Kim 2000).  Yamaguchi et al. (2001a,b) showed
that some of the shells and other high-pressure sources trigger
star formation directly in the LMC. Gouliermis et al. (2003) found
that stellar systems line the edges of the supershells found by
Kim et al. (1999). Thus some of the stellar hierarchy could be the
result of high-pressure triggering in hierarchical shells.

The connection between the hierarchical structures of the gas and
stars is emphasized further by the power spectra of optical
emission from galaxies. Elmegreen, et al. (2003a,b) found that
power spectra of optical emission along azimuthal cuts through
several galaxies have the same near-power law form as the power
spectra of HI emission from the LMC. In one case, M81, a spiral
density wave contributes to the power spectrum at the lowest
wavenumber, but otherwise the power spectrum is the same as in a
flocculent galaxy.

We might summarize these observations as follows: ``Star
complexes'' are the largest globular regions of star formation in
galaxies. They include associations of Cepheid variables, red
supergiants, WR stars, HII regions, and OB-associations (e.g.,
Ivanov 2004). They could be the origin of moving stellar groups
(Asiain et al. 1999). They are part of a continuum of
star-formation scales between clusters and swing-amplified
spirals. This continuum has at least two characteristics of
turbulence: power-law power spectra and a velocity-size relation.
We would like to know how star complexes form and how stars form
in them. Is there any evidence also for star complexes with a
characteristic scale, rather than an observationally selected
scale among many scales?

\section{Formation of Star Complexes}
Most galaxies with star formation have at least a few regions with
sizes comparable to the main stellar structures -- spiral arms,
tidal arms, resonance rings, etc..  Comparable sizes means
comparable to the minor dimensions, e.g., the widths of the arms
or rings. These regions are star complexes. They are probably the
largest scale in a local hierarchy of scales beginning with some
instability length and extending down to OB associations and
clusters. If this is the case, then the complexes will have a
characteristic length and mass. Galaxies without stellar spiral
waves or rings would not have a characteristic scale limited by
these structures, so they could produce an even wider range of
scales in the hierarchy of star formation. For example, the
optical structures could range from flocculent arms, which are
driven primarily by sheared gravitational instabilities, down to
star complexes, OB associations, and clusters, which are fragments
produced by self-gravity and turbulence (e.g., Huber \& Pfenniger
2001).

Recent CO observation of M33 (Engargiola et al. 2004), combined
with stellar complex data (Ivanov 2004) and older HI observations
(Deul \& van der Hulst 1987) show star complexes associated with
molecular and atomic gas. They also show that most molecular
clouds are inside giant HI clouds, as observed locally (Grabelsky
et al. 1987; Elmegreen \& Elmegreen 1987) and in other galaxies
(Lada et al. 1988). In addition, in M33, the CO cloud spin axes
are correlated up to scales of $\sim1$ kpc, suggesting coherence
on this scale. This is also the scale of the giant HI clouds. Thus
star formation proceeds first by forming giant HI clouds, and then
by forming molecular clouds inside of them and OB associations
inside the molecular clouds. Each collection of OB associations,
aged by $\sim30-100$ My, is a star complex (Efremov 1995). The
Gould's Belt region (e.g., Lallement et al. 2003) may be an
example. The regular distribution of giant HI clouds along spiral
arms has been known for many years, starting with the first HI
observations of the Milky Way (McGee \& Milton 1964) and
proceeding through the 1970's and 1980's when HI was routinely
mapped in nearby galaxies (e.g., Boulanger \& Viallefond 1992).
Regularity implies a characteristic scale, which is most likely
the Jean length (Elmegreen \& Elmegreen 1983; Kuno et al. 1995).

These observations demonstrate that star complexes form in CO/HI
cloud complexes. The stellar parts are hierarchically clumped, but
still coherent up to $\sim1$ kpc in the main disks of galaxies,
while the gaseous parts can extend for two or three times this
distance. The primary objects formed by galactic processes are
$\sim10^7$ M$_\odot$ HI clouds, while giant molecular clouds
(GMCs) are their fragments.  This $10^7$ M$_\odot$ mass is the
Jeans mass in the ambient ISM, suggesting that the HI clouds and
ultimately the star complexes form by gravitational instabilities.
The formation of GMCs follows by a combination of
self-gravitational contraction and turbulence compression inside
the $10^7$ M$_\odot$ clouds, but GMCs are not special, distinct
objects. The CO/HI ratio depends primarily on self-shielding, not
cloud formation processes. Low pressures, low metallicities, or
high radiation fields imply low CO/HI ratios in each cloud. CO/HI
varies with galactic radius or galaxy type because of variations
in pressure, radiation field and metallicity, without any change
in physical cloud structure or star formation properties
(Elmegreen \& Elmegreen 1987; Elmegreen 1993; Honma, Sofue, \&
Arimoto 1995; Engargiola et al. 2004).

There are 2 basic dynamical phases and 2 basic chemical phases for
neutral clouds: atomic or molecular gas in a self-gravitating
cloud, and atomic or molecular gas in a non-self-gravitating
cloud. Usually the atomic phase dominates at low density, the
molecular phase at high density, the non-self-gravitating phase at
low column density, and the self-gravitating phase at high column
density (Elmegreen 1995). The virial theorem also plays a role:
considering the presence of some external pressure, both diffuse
and self-gravitating clouds occur at low mass but only self
gravitating clouds occur  at high mass.  In the general ISM,
gravitating cloud complexes are evident mostly from knots in
spiral arms, spurs, and the presence of star formation (e.g., Kim
\& Ostriker 2002).  In accord with the virial theorem result
above, the FCRAO Outer Galaxy Survey (Heyer, Carpenter \& Snell
2001) shows that self-gravity is important only in the most
massive CO clouds, $M>10^4$ M$_\odot$. This is the same mass at
which the virial theorem suggests a transition from both diffuse
and self-gravitating clouds (at lower mass) to purely
self-gravitating clouds (at higher mass), given a near-constant
diffuse cloud density of $\sim50$ cm$^{-3}$ (Elmegreen 1995). This
mass limit should depend on pressure and the molecule detection
threshold. Small self-gravitating cores are undoubtedly present
inside these FCRAO clouds (because stars are forming), but they
have a higher density threshold for detection and a smaller
angular size, making their detection not as likely in large-scale
CO surveys.

\section{Characteristic Size versus Scale-Free?} If a galaxy has
a global spiral density wave in the stars, or if it has a stellar
ring, then gaseous gravitational instabilities in these structures
have a size and mass defined by the stellar geometry: i.e., the
instability length is $\sim3\times$ the spiral arm (or ring)
width. Examples are the well-known ``beads on a string of star
formation'' along spiral arms,  and the ``nuclear ring hotspots.''
If a galaxy has no spiral density wave, then the stars and gas
become unstable together, forming multiple spiral arms or
flocculent arms that are made of old stars, gas and star
formation. The instability involved is the swing amplifier (Toomre
1981), usually enhanced by magnetic fields (Kim, Ostriker \& Stone
2002, 2003). The instability should also drive turbulence,
producing scale-free clouds and star formation as observed.

Stellar spirals define two characteristic scales: $2\pi
G\Sigma/\kappa^2$, which is the Toomre (1964) length for the
separation between spiral arms, and $2c^2/G\Sigma$, which is the
Jeans length for pressure balance against self-gravity. Here,
$\kappa$ is the epicyclic frequency, $\Sigma$ is the mass column
density, and $c$ is the gas velocity dispersion. The Jeans mass is
$c^4/G^2\Sigma$. Inside spiral arms, the Jeans instability is
one-dimensional, i.e., the collapse is parallel to the arms, and
the characteristic length is usually about 3 times the arm width
(Elmegreen \& Elmegreen 1983; Bastien et al. 1991). The condition
for rapid instability in this case is not the Toomre $Q$
condition, but the 1-dimensional analog: $\pi G\mu/c^2>1$ for
mass/length $\mu$ (Elmegreen 1994).

Wada \& Norman (2001) modeled 2D hydrodynamics of galaxy disks
without spiral density waves and found that gravitational
instabilities drive turbulence, giving a log-normal density pdf
that is typical for isothermal compressible turbulence
(Vazquez-Semadeni 1994). Three-dimensional SPH models of galaxy
disks without spiral density waves get the Schmidt/Kennicutt laws
of star formation (Li et al. 2004).

\section{Theory of the Star Formation Rate} A sensible local SF
law is (Elmegreen 2002b)
\begin{equation}{\rm SFR/V} = \epsilon \rho (G\rho)^{1/2},
\label{eq1}
\end{equation}
which is the efficiency times the mass per unit volume, times the
conversion rate from gas into stars. Kennicutt (1998) observes
\begin{equation}{\rm SFR/Area} = 2.5\times10^{-4}
\left(\Sigma/{\rm M}_\odot\; {\rm pc}^{-2}\right)^{1.4}\;\;{\rm
M}_\odot\;{\rm kpc}^{-2}\;{\rm yr}^{-1}\sim
0.033\Sigma\Omega\end{equation} for average mass column density
$\Sigma$ in the whole disk and rotation rate in the outer part
$\Omega$.

If we convert the global SFR/A into a local SFR/V using the local
tidal density for gas, $\rho =3\Omega^2/2\pi G$, a flat rotation
curve, and an exponential disk with scale length
$r_D/r_{edge}=0.25$, then the average SFR/A converts to a local
\begin{equation}{\rm SFR/V}=0.012 \rho(G\rho)^{1/2}.
\label{sf}
\end{equation}

Why is the efficiency $\epsilon= 0.012$, and is this the right
star formation law? Boissier et al. (2003) compared the star
formation rates versus radii in 16 galaxies with three simple
expressions, finding factor of 3 variations around each law with
no apparent cause, and no preferred law.  Either we do not know
the ``right'' star formation law, or additional processes give big
variations around one of the assumed laws.  By the way, all of the
Boissier et al. laws, and that discussed by Hunter, Elmegreen, \&
Baker (1998) for dwarf irregular galaxies, are consistent with a
local star formation rate proportional to the stellar surface
density. Whether this is a cause or an effect of star formation is
not clear. That is, one might expect the star formation rate to
scale with the existing stellar surface density if star formation
is building up that surface density and the exponential scale
length does not change much with time. On the other hand, the
relationship might also exist if background stars trigger star
formation, as might be the case if supernova and HII regions
directly compress the gas to trigger star formation or if these
pressures indirectly compress the gas by driving supersonic
turbulence.

Let us proceed with equation (\ref{sf}) and ask what determines
the star formation efficiency, $\epsilon$.  Assume that stars form
in dense cores where the efficiency is $\epsilon_c\sim0.5$ and
$\rho_c\sim10^5$ cm$^{-3}$, giving
\begin{equation}
{\rm SFR/V}_{core} =
\epsilon_c\rho_c\left(G\rho_c\right)^{1/2}.\label{eq4}\end{equation}
Then make a conversion:
\begin{equation} {\rm SFR/V}_{gal} = {\rm SFR/V}_{core} \times
\left(V_{core}/V_{gal}\right)={\rm SFR/V}_{core}\times
\left(M_{core}/M_{gal}\right)\times\rho/\rho_c,\end{equation}
where $\rho_c=M_{core}/V_{core}$, $\rho=M_{gal}/V_{gal}$,
$M_{core}$ and $V_{core}$ are the summed mass and volume of all
cores, while $M_{gal}$ and $V_{gal}$ are the total gas mass and
gas volume of the galaxy. Now substitute from equation \ref{eq4}
and set ${\rm SFR/V}_{gal}$ equal to the observed rate $0.012
\rho(G\rho)^{1/2}$. Then we get the gas mass fraction in
star-forming cores:
\begin{equation}f_M\equiv\left(M_{core}/M_{gal}\right)\equiv
0.012/\epsilon_c\times\left(\rho/\rho_c\right)^{1/2} \sim
10^{-4}.\end{equation} Thus the observed average efficiency of
$\epsilon= 0.012$ (per dynamical time) requires $10^{-4}$ of the
total ISM mass to be in star-forming cores if all regions evolve
on a dynamical timescale (Elmegreen 2002b). This is the same mass
fraction as in the Wada \& Normal (2001) log-normal for
$\rho/\rho_{ave}
> 10^5$, as assumed above with $\rho_c\sim10^5$ cm$^{-3}$ and
$\rho\sim1$ cm$^{-3}$. This result depends on the density pdf for
the ISM, which is not observed yet, and it assumes a log-normal
form for this pdf.  In fact, the high density portion may become a
power-law after collapse starts (Klessen 2000). Nevertheless, the
agreement between the simple theory and the Kennicutt (1998) star
formation rate, which applies to essentially all late-type galaxy
disks and their nuclear regions, suggests that something universal
like turbulence helps partition the gas in a hierarchical fashion
and that only the dense regions at the bottom of the hierarchy
form stars.

A turbulent ISM has a small fraction of its mass at a high enough
density to form stars.  Most of the mass is either too low a
density to form stars, or is not self-gravitating enough to resist
turbulent disruption.  For a turbulent medium, every structure
forms on a local crossing time, but the progression to high
density is not monotonic. The low-density clumps are destroyed
easily and they are smashed and sheared into smaller pieces by
transient pressure bursts. The progress toward high density is
more like a random walk, with some interactions making denser
regions and some making lower densities. Eventually the lucky ones
that had a long succession of compressive interactions become
dense enough and massive enough to be strongly self-gravitating at
the typical pressure in the cloud. Then they presumably produce
stars quickly. The delays from magnetic diffusion, disk formation,
and turbulent energy dissipation are not nearly as time consuming
as the fragmentation process on larger scales. Thus the largest
scales control the overall rate.  All that a microscopic delay
might do is change the form of the density pdf, producing a bump
at high density, for example, or a power-law instead of a
log-normal, if the collapse slows down. If only turbulence is
involved, though, the random walk in density produces a log-normal
density pdf (Vazquez-Semadeni 1994).  Our integration over the pdf
for densities $\rho_c/\rho>10^5$ involves an assumption that the
collapse delays occur at higher densities, where the detailed
shape of the pdf will not affect the integral under it if there is
a steady flow toward higher density during the star formation
process.

If we now denote the galactic average quantities by a subscript
``0'', then the efficiency at any density is given by:
\begin{equation}\epsilon_0 \rho_0 (G\rho_0)^{1/2} =
\epsilon(\rho) \rho (G\rho)^{1/2}f_V(\rho) = \epsilon_c
\rho_c(G\rho_c)^{1/2}f_V(\rho_c).\end{equation} But $\rho
f_V(\rho) = \rho_0f_M(\rho)$, etc., so
\begin{equation}\epsilon(\rho) = \epsilon_c
\left(\rho_c/\rho\right)^{1/2} \left[f_M\left(\rho_c\right)/
f_M\left(\rho\right)\right].\end{equation} Here, $f_V(\rho)$ is
the fraction of the volume having a density larger than $\rho$ and
$f_M(\rho)$ is the fraction of the mass having a density larger
than $\rho$.

\begin{figure}[ht]
\centerline{\includegraphics[width=3in]{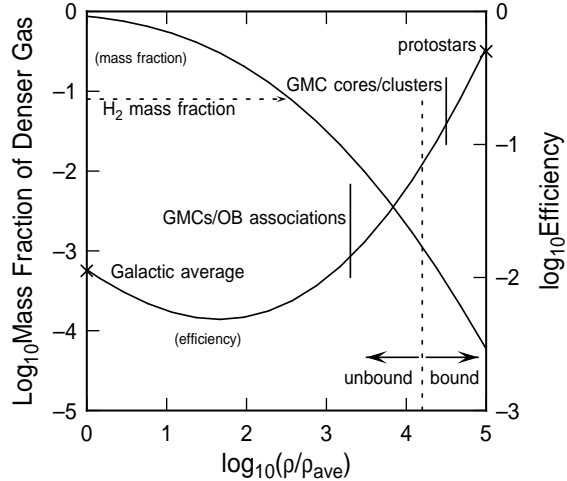}}\caption{The
mass fraction (downward-sloping curve) and efficiency of star
formation as a function of density, derived from the log-normal
density pdf in Wada \& Norman (2001).  Observed efficiencies in
cloud cores and OB associations are shown as vertical lines. The
increase of efficiency with density suggests that star formation
proceeds in a hierarchical medium.}
\end{figure}

Figure 1 shows as a decreasing line the mass fraction,
$f_M(\rho)$, versus density $\rho$ using the log-normal found by
Wada \& Norman (2001) for a 2D disk with star formation,
turbulence, and self-gravity. The scale for $f_M$ is on the
left-hand axis. The figure also shows as an increasing line the
efficiency $\epsilon(\rho)$, using the right-hand axis.  The
efficiency increases with increasing average density because the
hierarchical nature of clouds gives them a higher filling factor
for dense gas at higher average density. The mass fraction
decreases with density because only a small fraction of the matter
is dense. Several commonly observed values for the efficiency are
indicated: the average galactic value of 0.012, derived above, the
range of $\sim1-5$\% for whole OB associations, and the range of
$\sim10-30$\% for the cores of OB associations, where most stars
actually form, producing clusters. Note that this efficiency is
not the local efficiency where single stars form; that is assumed
to be the constant value of $\epsilon_c=0.5$ at
$\rho/\rho_{ave}=10^5$. Rather, it is the efficiency inside a
cloud whose boundary density is $\rho$. OB associations form with
an overall low efficiency because there is a lot of inactive gas
at low density. Only the cores, and in them, only the small dense
cores of these cores, form stars with high efficiency. This
increase of efficiency with cloud density is commonly observed and
easily explained for hierarchical stellar regions when all star
formation occurs locally at the highest density.  Note that star
formation in simulations (e.g., Mac Low \& Klessen 2004) proceeds
over some prolonged time as does dense core formation, but the
total time for this is still about the crossing time on the
largest scale, making equation \ref{eq1} appropriate.

Figure 1 highlights the boundary between bound and unbound stellar
regions, which is where the average efficiency is greater than
several tenths (Lada \& Lada 2003).  Bound regions have high
densities and may therefore be identified with clusters rather
than OB associations or any other part of the hierarchy on larger
scales. The masses of the bound regions are not specified by this
derivation but may be anything, always distributed as $dN/dM\sim
M^{-2}$ for hierarchical gas structures (Fleck 1996; Elmegreen \&
Efremov 1997). This is the essential explanation for the formation
of most stars in clusters. After $\sim10$ My, most clusters
dissolve and their stars fill in the region where they once
clustered together. This is an OB association. In another
$\sim30-100$ My, these OB associations dissolve and fill in the
region where they clustered together;  this makes a star complex.
Star complexes are so big that their dispersal is relatively slow
and therefore accompanied by significant shear in normal galaxy
disks. Thus there is no globular-shaped super-collection of star
complexes, only flocculent spiral arms on larger scales.

Heyer et al. (2004) studied the star formation rate versus radius
for molecular gas in M33. Both the SFR and the column density
increase toward the center, with a mutual relationship
\begin{equation} {\rm SFR} = 3.2 \left(\Sigma_{H2}/
{\rm M}_\odot\; {\rm pc}^{-2}\right)^{1.36} {\rm M}_\odot {\rm
pc}^{-2}\;{\rm Gy}^{-1}. \end{equation} This may be written as in
eq. (\ref{eq1}),  ${\rm SFR}\sim 0.6 \rho_{H2}
(G\rho_{H2})^{1/2}$, assuming a disk thickness of 150 pc. The
efficiency is higher for the distributed CO density
($\epsilon_{H2}\sim0.6$) than it is for the total gas density
($\epsilon_0\sim0.012$) because the average density of $H_2$
spread around a disk is lower than the total density by
$\rho_{H2}/\rho_0=\left(\epsilon_0/\epsilon_{H2}\right)^{2/3}=
\left(0.012/0.6\right)^{2/3} = 0.08$ (i.e., 0.6 is not the
$\epsilon$ inside a CO cloud). This is the $M_{H2}/M_{gal}$ mass
ratio actually obtained from Figure 1 if the density of
CO-emitting material is taken to be $\sim300$ cm$^{-3}$ (use the
left-hand axis). Thus the CO-Schmidt law follows from this
hierarchical model too.

As mentioned briefly above, giant molecular clouds, OB
associations, and star clusters all get their mass distribution
functions from the structure of a compressibly turbulent medium.
These mass functions are a property of fractals, sampled in
various ways (Elmegreen 2002a; also see Elmegreen 2004).  For a
region sampled at low density, far from the peak, the mass
function is approximately a power law with a shallow slope,
$\sim-1.5$, as observed for GMCs. When the same region is sampled
at a higher density, the mass function has a steeper slope,
$\sim-2$, as observed for clusters. This explains how GMCs and
clusters can both form from the same gas distribution and yet have
slightly different mass functions.  Note that this works because
the the ratio of cluster mass to cloud-core mass is about constant
when there is a threshold efficiency required for bound cluster
formation.

\section{Conclusions}

Star complexes usually form in clouds that result from a
gravitational instability in the ISM.  If there is an imposed
structure on the gas distribution, such as a stellar spiral arm or
ring, then the clouds can line up along this structure with a
semi-regular spacing, producing beads on a string in spiral arms,
tidal arm star-forming clumps, collisional ring star formation,
nuclear ring hotspots, dwarf galaxy hot spots, etc.. The cloud
typically has $10^7$ M$_\odot$ of gas in the main galaxy disk, and
it makes a $10^5$ M$_\odot$ star complex over a $\sim50-100$ My
period. In galaxies with no imposed stellar structures, gaseous
instabilities and turbulence compression still make giant clouds
and their fragments, but the star complexes are usually selected
to be the largest globular scale, excluding the flocculent arms
themselves. In both cases, star complexes appear as groupings of
intermediate-age stars, such as Cepheid variables and red
supergiants.

For a turbulent self-gravitating medium, star formation should
operate on about the local dynamical time over a wide range of
scales. Turbulence structures the gas, placing only a small
fraction of the mass at a high enough density to form stars. The
resulting hierarchical structure is somewhat continuous from the
scale of the disk thickness to individual pre-stellar
condensations. Stars form in the turbulent structures, making a
hierarchy of stellar complexes, associations, clusters, multiple
stars, and binaries. The efficiency of star formation increases
with density in such a hierarchical medium. Only high density
regions have high enough efficiencies to form bound clusters.

\begin{acknowledgments}
Much of this work was done with the help of a National Science
Foundation grant AST-0205097.
\end{acknowledgments}

\end{document}